\documentclass[usenatbib]{mn2e}
\usepackage{graphicx}
\usepackage{stfloats}
\usepackage{amsmath}
\usepackage{amssymb}
\usepackage{color}

\title[Implications of the Eccentric Kozai-Lidov Mechanism]{Implications of the Eccentric Kozai-Lidov Mechanism for Stars Surrounding Supermassive Black Hole Binaries}

\author[Li, Naoz, Kocsis and Loeb]{Gongjie Li$^1$, Smadar Naoz$^2$, Bence Kocsis$^3$, Abraham Loeb$^1$ \\
$^1$ Harvard-Smithsonian Center for Astrophysics, The Institute for Theory and
Computation, 60 Garden Street, Cambridge, MA 02138, USA \\
$^2$ Department of Physics \& Astronomy, Division of Astronomy and Astrophysics, UCLA, Los Angeles, CA 90095, USA \\
$^3$ Institute for Advanced Study, Princeton, NJ 08540, USA }

\begin{document}
\bibliographystyle{mn2e}
\maketitle

\newcommand{\apj}{ApJ}
\newcommand{\apjl}{ApJL}
\newcommand{\apjs}{ApJS}
\newcommand{\mnras}{MNRAS}
\newcommand{\aap}{AAP}
\newcommand{\prd}{PRD}
\newcommand{\aj}{AJ}
\newcommand{\pasp}{PASP}
\newcommand{\araa}{ARA\&A}
\newcommand{\nat}{Nature}
\newcommand{\na}{New Astron.}
\newcommand{\nar}{New Astron. Rev}
\newcommand{\apss}{Ap\&SS}
\newcommand{\actaa}{Acta Astron}
\newcommand{\planss}{Planetary Space Sci.}
\newcommand{\icarus}{Icarus}
\newcommand{\be}{\begin{equation}}
\newcommand{\ee}{\end{equation}}
\newcommand{\bea}{\begin{eqnarray}}
\newcommand{\eea}{\end{eqnarray}}
\newcommand{\BK}[1]{{\color{red}\bf BK: {#1}}}
\newcommand{\GJ}[1]{{\color{blue}\bf GJ: {#1}}}

\def\Mpc{\rm Mpc}
\def\Mbh{M_{\rm BH}}

\def\Msun{{\rm M}_{\odot}}
\def\kpc{\rm kpc}
\newcommand{\comment}[1]{}



\begin{abstract}
An enhanced rate of stellar tidal disruption events (TDEs) may be an important characteristic of supermassive black hole (SMBH) binaries at close separations. Here we study the evolution of the distribution of stars around a SMBH binary due to the eccentric Kozai-Lidov (EKL) mechanism, including octupole effects and apsidal precession caused by the stellar mass distribution and general relativity. We identify a region around one of the SMBHs in the binary where the EKL mechanism drives stars to high eccentricities, which ultimately causes the stars to either scatter off the second SMBH or get disrupted. For SMBH masses $10^7 M_{\odot}$ and $10^8 M_{\odot}$, the TDE rate can reach $\sim10^{-2}/$yr and deplete a region of the stellar cusp around the secondary SMBH in $\sim0.5$ Myr. As a result, the final geometry of the stellar distribution between 0.01 and 0.1 pc around the secondary SMBH is a torus. These effects may be even more prominent in nuclear stellar clusters hosting a supermassive and an intermediate mass black hole.
\end{abstract}



\section{Introduction}
\label{s:intro}
Supermassive black holes (SMBHs) are ubiquitous at the centers of galaxies \citep{Kormendy13}. Stars passing close to the SMBH can be tidally disrupted, and the fall back of the stellar debris produces a strong electromagnetic tidal disruption flare \citep[e.g.,][]{Gezari12R}. More than a dozen tidal disruption event (TDE) candidates have been observed until present \citep[e.g.,][]{Bade96, Gezari03, Gezari06, Gezari08, Gezari09, vanVelzen11, Gezari12, Holoien14}, including two candidates with relativistic jets \citep{Levan11, Bloom11, Zauderer11, Cenko12}. TDEs can provide valuable information on dormant SMBHs, which are otherwise difficult to detect.

The rate of the TDEs provide information about the SMBH and the stellar distribution in the center of galaxies \citep{Stone14}. The rate of TDEs is highly uncertain observationally due to the small sample size. It is estimated to be in the range of $10^{-5} - 10^{-4}$ per galaxy per year by \citet{Donley02, Gezari08, Maksym12, vanVelzen14}. This roughly agrees with the theoretical estimates, discussed by \citet{Frank76, Lightman77, Cohn78, Magorrian99, Wang04, Brockamp11, Stone14}. However, the TDE rate may be enhanced due to the presence of a non-axisymmetric gravitational potential around the SMBH \citep{Merritt04}, or due to a massive perturber \citep{Perets07}. In addition, the TDE rate may be higher in galaxies with more than one SMBH \citep{Ivanov04, Chen09, Chen11, Wegg11}, or when the SMBH binary (SMBHB) recoils due to the emission of gravitational waves \citep{Stone11, Li12, Stone12}. Some TDEs may not appear as flares and therefore be missed in observations\citep{Guillochon15}.

The interaction between a SMBHB and an ambient star cluster has been discussed in the literature using numerical scattering experiments by \citet{Sesana11} and using direct N-body simulations by \citet{Iwasawa11, Gualandris11, Meiron13, Wang14}. In particular, it has been shown that the star cluster may either increase or decrease the eccentricity of the SMBHB depending on the fraction of counter-rotating to co-rotating stars. The SMBHB ejects a population of stars from the cluster in an anisotropic manner, and the SMBHB produces a deficit in the number density of stars, a dip in the velocity dispersion in the inner regions, and an inner counter-rotating and an outer co-rotating torus of stars with respect to the binary.

In this paper, we focus on the distribution of stars orbiting close to one member of the SMBHB and perturbed by the other SMBH through hierarchical three-body interactions. We examine the effect of these hierarchical three body interactions. Specifically, the outer SMBH perturbs the stellar population around the inner\footnote{We consider stars that initially orbit the ``inner'' SMBH and whose orbits are perturbed by the ``outer'' SMBH regardless of which SMBH is more massive.} SMBH, and leads to long-term variations in the eccentricities and inclinations of the stellar orbits while keeping the semimajor axes of their orbits fixed. In particular, when the orbit of the SMBH secondary is circular and if the mutual inclination between the orbits of the SMBHB and a star is over $40^\circ$, the stellar eccentricity and inclination undergo periodic oscillations, known as the quadrupole Kozai-Lidov mechanism \citep{Kozai62, Lidov62}. This is caused by the long-term (orbit-averaged) Newtonian gravitational effect expanded in multipoles to the quadrupole order, i.e. second order in the semimajor axis ratio of the stellar and the outer SMBH's orbit. More generally, it has been found that when the outer orbit is eccentric, the analogous octupole eccentric Kozai-Lidov mechanism (EKL, third order in semi major axis ratio) causes the eccentricity to be excited very close to unity and the inner orbit to flip relative to the invariable plane from prograde to retrograde or vice versa \citep[e.g.,][]{Ford00, Naoz11a, Katz11, Lithwick11, Naoz13a, Naoz13b, Li14a, Li14b}. The TDE rate has been discussed in the literature for stars orbiting an SMBHB, where the quadrupole Kozai-Lidov mechanism can enhance the tidal disruption rate \citep{Ivanov04, Chen09, Chen11, Wegg11}. For the Galactic Center, the Kozai-Lidov mechanism driven by the stellar disk has also been discussed and the additional effects of Newtonian apsidal precession were shown to play a significant role \citep{Chang09}. In light of recent developments in the understanding of hierarchical three body interactions 
we revisit this problem. Since the stellar eccentricity can be increased to a value much closer to unity by eccentric perturbers, we expect the EKL mechanism to enhance TDE rates with respect to the circular case. We therefore seek to re-evaluate the total number of stars vulnerable to TDE due to EKL.

It is well known that apsidal precession quenches the EKL mechanism \citep[e.g.,][]{Ford00,Blaes02, Naoz13b}. In galactic nuclei this may be due to the Newtonian (NT) gravitational effect of the spherical stellar cusp or general relativistic (GR) precession, provided that the corresponding precession timescale is much shorter than the Kozai timescale \citep{Chang09}. Furthermore, the EKL mechanism may be quenched if the eccentricity of the star is changed by the stellar cluster due to scalar resonant relaxation, or if the orbital plane is reoriented by the stellar cluster due to vector resonant relaxation \citep{Rauch96, Kocsis11, Kocsis14} or Lense-Thirring precession \citep{Merritt10,Merritt12}. We find that NT precession and GR precession may have a large effect on the EKL mechanism, but tidal effects, scalar and vector resonant relaxation, and Lense-Thirring precession are typically less important. The timescale on which the EKL mechanism operates increases if the outer SMBH mass is reduced. Thus, GR 
precession may dominate over and quench the EKL mechanism most efficiently if the outer SMBH is less massive than the inner SMBH (see figure 2 in \citealt{Naoz14}). Similarly, we find that NT precession also suppresses the EKL mechanism most efficiently when the outer SMBH is less massive. Tidal disruption is expected in the opposite regime when the EKL mechanism is very prominent, i.e. when the outer SMBH is more massive than the inner SMBH. We identify the outcome of the EKL mechanism as a function of SMBHB parameters and quantify the TDE rate.

Our discussion is organized as follows. In \textsection 2, we describe the adopted methods. In \textsection 3, we characterize the parameter space to identify where the EKL mechanism is important. Then, we calculate the tidal disruption rate and discuss the final stellar distribution due to the EKL mechanism with an illustrative example in \textsection 4, and for stars surrounding an intermediate-mass black hole in \textsection 5.  Finally, we summarize our main results in \textsection 6.

\section{Method}
\label{s:Method}
\begin{figure}
\includegraphics[width=3.2in,height=2.1in]{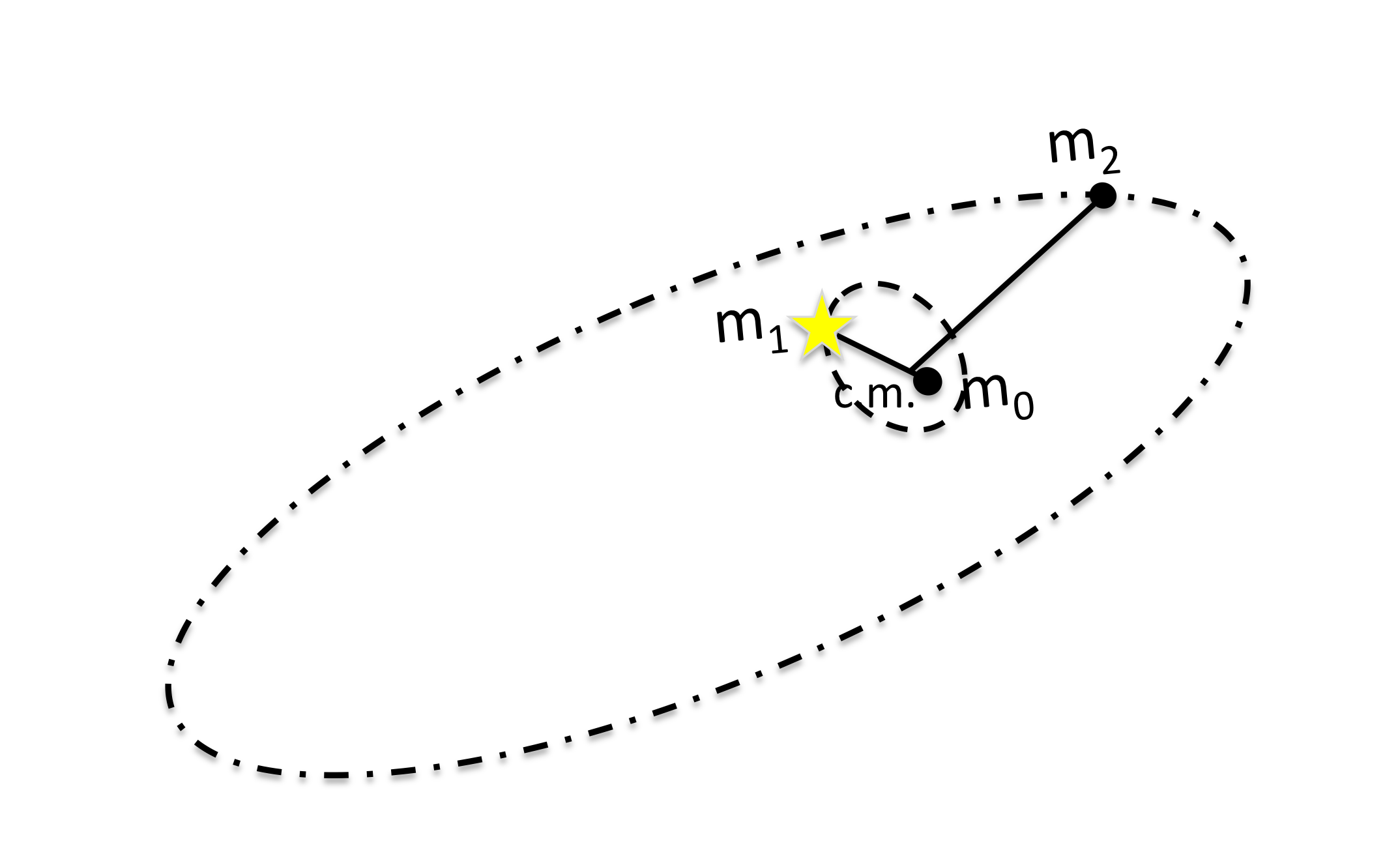}
\caption{\label{f:config} The system configuration. `c.m.' denotes the center of mass of the inner binary, which contains the star (with mass $m_1$) and SMBH (with mass $m_0$). The other SMBH (with mass $m_2$) is on an outer orbit.}
\vspace{0.1cm}
\end{figure}

We study the tidal disruption of stars due to the EKL mechanism in galaxies that host a SMBHB. The three-body system consists of an ``inner binary'' comprised of the SMBH and a star, and an ``outer binary'' comprised of the outer SMBH and the center mass of the inner binary, as shown in Figure \ref{f:config}. We denote the masses of the objects by $m_0$ (inner SMBH), $m_1$ (star), and $m_2$ (outer SMBH), and for orbital parameters we use subscript 1 and 2 for the inner and outer binary, respectively. In order for the EKL mechanism to operate, we require the triple system to be in a hierarchical configuration: the inner binary on a much tighter orbit than the third object, such that \citep[e.g.,][]{Lithwick11, Katz11},
\begin{equation}
\epsilon = \frac{a_1}{a_2} \frac{e_2}{1-e_2^2} < 0.1 \, ,
\label{eqn:eps} 
\end{equation}
where $a$ and $e$ are respectively the semimajor axis and eccentricity.

\subsection{Comparison of timescales}
\label{s:Cot}
We examine the range of orbital parameters in oder to identify the regions in which the EKL mechanism may operate. 
The relevant processes' timescales can be expressed as:
\begin{align}
t_{K} &= \frac{2\pi a_2^3 (1-e_2^2)^{3/2} \sqrt{(m_0+m_1)(1-e_1^2)}}{\sqrt{G}a_1^{3/2}m_2} \\
t_{oct} &= \frac{1}{\epsilon} t_{K}\\
t_{GR1} &= \frac{2 \pi a_1^{5/2} c^2 (1-e_1^2)}{3G^{3/2}(m_0+m_1)^{3/2}} 
\end{align}\begin{align}
t_{GR2} &= \frac{2\pi a_2^{5/2} c^2 (1-e_2^2)}{3G^{3/2}(m_0+m_1+m_2)^{3/2}} \\
t_{GR, int} &=\frac{16}{9}\frac{a_2^3 c^2 (1 - e_2^2)^{3/2}(m_0)^{3/2}}{\sqrt{a_1}e_1\sqrt{1 - e_1^2}G^{3/2}m_0^2m_2}\\
t_{NT} &= 2\pi \Big(\frac {\sqrt{Gm_0/a_1^3}}{\pi m_0 e_1} \int_0^\pi d\psi \, M_* (r) \cos{\psi} \Big)^{-1} \\
t_{RR, s} &= \frac{4\pi \omega}{\beta_s^2\Omega^2}\frac{m_0^2}{M_*(r)m_1} \\
t_{RR, v} &= \frac{2\pi f_{vrr}}{\Omega}m_0\frac{1}{\sqrt{M_*(r)m_1}}\\
t_{rel} &= 0.34\frac{\sigma^3}{G^2\rho m_1\ln{\Lambda}}\\
t_{LT} &= \frac{a_1^3c^3(1-e^2)^{3/2}}{2G^2m_0^2s}\\
t_{GW} &= \frac{a_2^4}{4}\frac{5}{64}\frac{c^5}{G^3m_0m_2(m_0+m_2)} \, .
\end{align}
Here $t_K$ is the quadrupole (${\cal O}(a_1 / a_2)^2$) Kozai timescale. Following \citet{Naoz13b}, $t_{oct}$ is the octupole (${\cal O}(a_1 / a_2)^3$) Kozai timescale. $t_{GR1}$ and $t_{GR2}$ are the timescales of the first order post Newtonian general relativistic (GR) precession at the quadrupole order (${\cal O}(a_1 / a_2)^2$) on the inner and outer orbit, and $t_{GR, int}$ is the timescale associated with the first post-Newtonian order GR interaction between the inner and the outer orbit. Following \citet{Kocsis11}, $t_{NT}$ is the timescale of the Newtonian precession caused by the stellar potential, and $t_{RR, s}$ and $t_{RR,v}$ are the timescales of the scalar and vector resonant relaxation. $t_{rel}$ is the two body relaxation timscale. $t_{LT}$ is the Lense-Thirring precession timescale, and $t_{GW}$ is the timescale of the orbital decay of the binary SMBHB due to gravitational wave radiation. For the resonant relaxation timescales, $M_* (r)$ is the mass of the 
stars interior to $r$, $\omega$ is the net rate of precession 
due to GR and NT, $\beta_s$ is estimated to be $1.05\pm0.02$ by \citet{Eilon09}, $\Omega$ is the orbital frequency of the star, and $f_{vrr}$ is estimated to be $1.2$ by \citet{Kocsis14}. For the Lense-Thirring timescale, $sGm_0^2/c$ is the spin angular momentum of the inner SMBH \citep[see references in e.g., ][]{Naoz13b, Kocsis11, Peters64}. We define some of these effects in more detail in \textsection~\ref{s:EOM} below.

The EKL mechanism operates if the following criteria are satisfied: 
\begin{enumerate}
 \item \label{i:hierarchical} The three-body configuration satisfies the hierarchical condition ($\epsilon < 0.1$, see equation (\ref{eqn:eps})) 
 \item The stars stay in the Hill sphere of the inner SMBH in order for them to remain bound to it, i.e. $a_1(1+e_1) < a_2(1-e_2)(m_0 / 3 m_2)^{1/3}$. 
 \item \label{i:quadrupole} The quadrupole (${\cal O} (a_1/a_2)^2$) Kozai timescale, $t_{K}$, needs to be shorter than the timescales of the other mechanisms that modify the orbital elements, otherwise the EKL mechanism is suppressed. The competing mechanisms include Newtonian precession (NT), general relativistic precession (GR), scalar resonant relaxation, vector resonant relaxation, two-body relaxation, Lense-Thirring precession, and the gravitational radiation.  
 \end{enumerate}

Note that the secular approximation fails when the perturbation from the outer SMBH is too strong or when the eccentricity reaches values very close to unity \citep[e.g.,][]{Antonini12, Katz12, Antognini13, Antonini14, Bode14}. This means that there are some systems that are poorly described by our approximation. However, we expect that those systems reach even higher eccentricities than the one predicted by the octupole approximation (e.g., Antognini et al. 2013), and thus our overall qualitative conclusions may hold even for those systems, but the quantitative rate values possibly underestimate the true rates. 

To calculate the Newtonian timescale, the resonant relaxation timescales, and the two body relaxation timescale, we adopt the spherically symmetric model for the stellar density discussed in \citet{OLeary09}. Specifically, the stellar density distribution is a power law of semimajor axis and the normalization is fixed by the $M-\sigma$ relation,
\begin{equation}
\label{e:rho}
\rho_*(r)=\frac{3 - \alpha}{2\pi} \frac{m_0} {r^3} \left(\frac{GM_0(m_0/M_0)^{1 - 2/k}}{\sigma_0^2 \, r}\right)^{-3 + \alpha} ,
\end{equation}
where $k = 4$, $M_0 = 1.3 \times10^8 \Msun$, $\sigma_0 = 200 {\rm km} / {\rm s} $ \citep{Tremaine02}, and we set $\alpha = 1.75$.

\begin{figure}
\includegraphics[width=2.9in, height=1.7in]{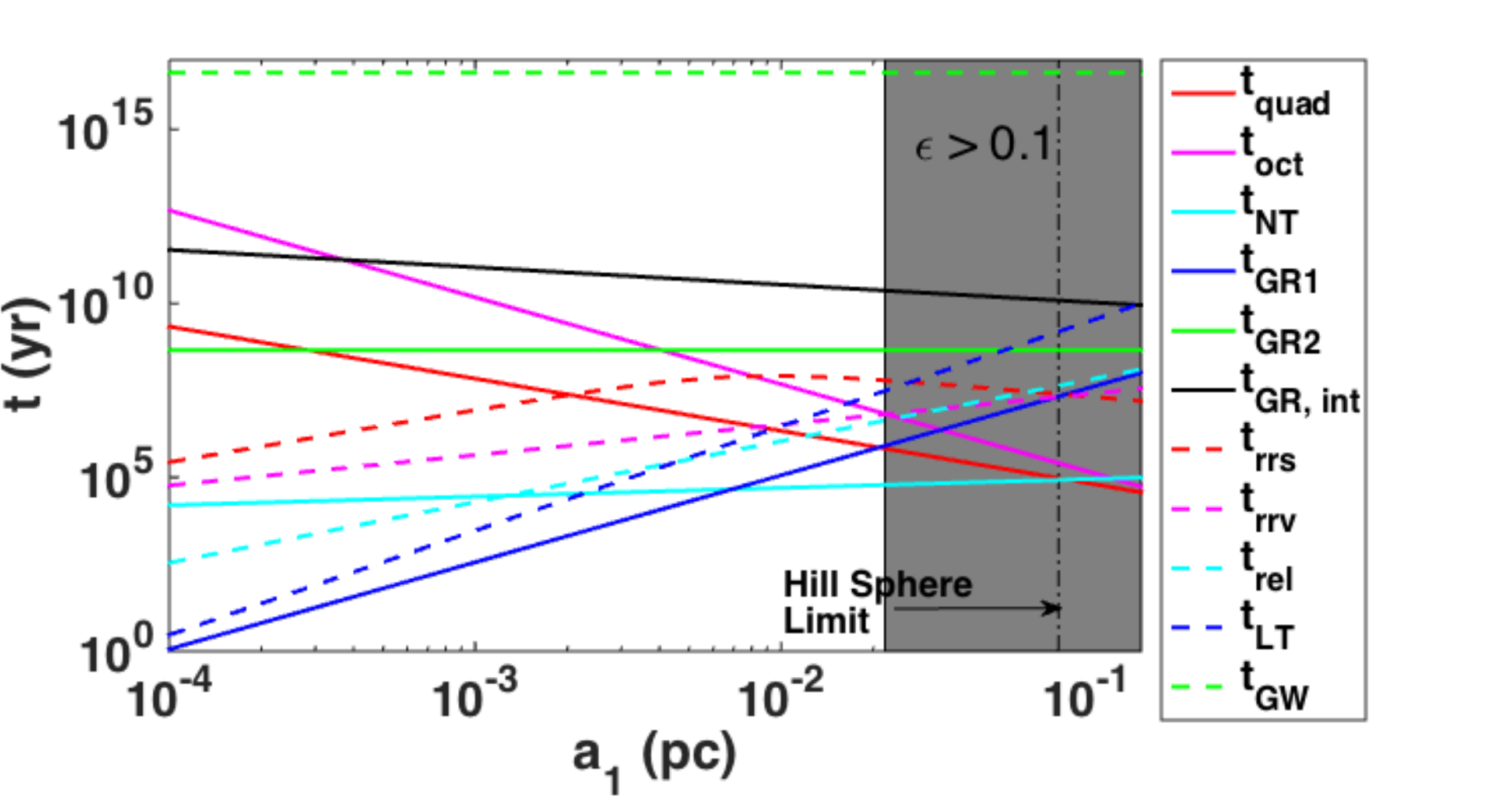}\\
\includegraphics[width=2.9in, height=1.7in]{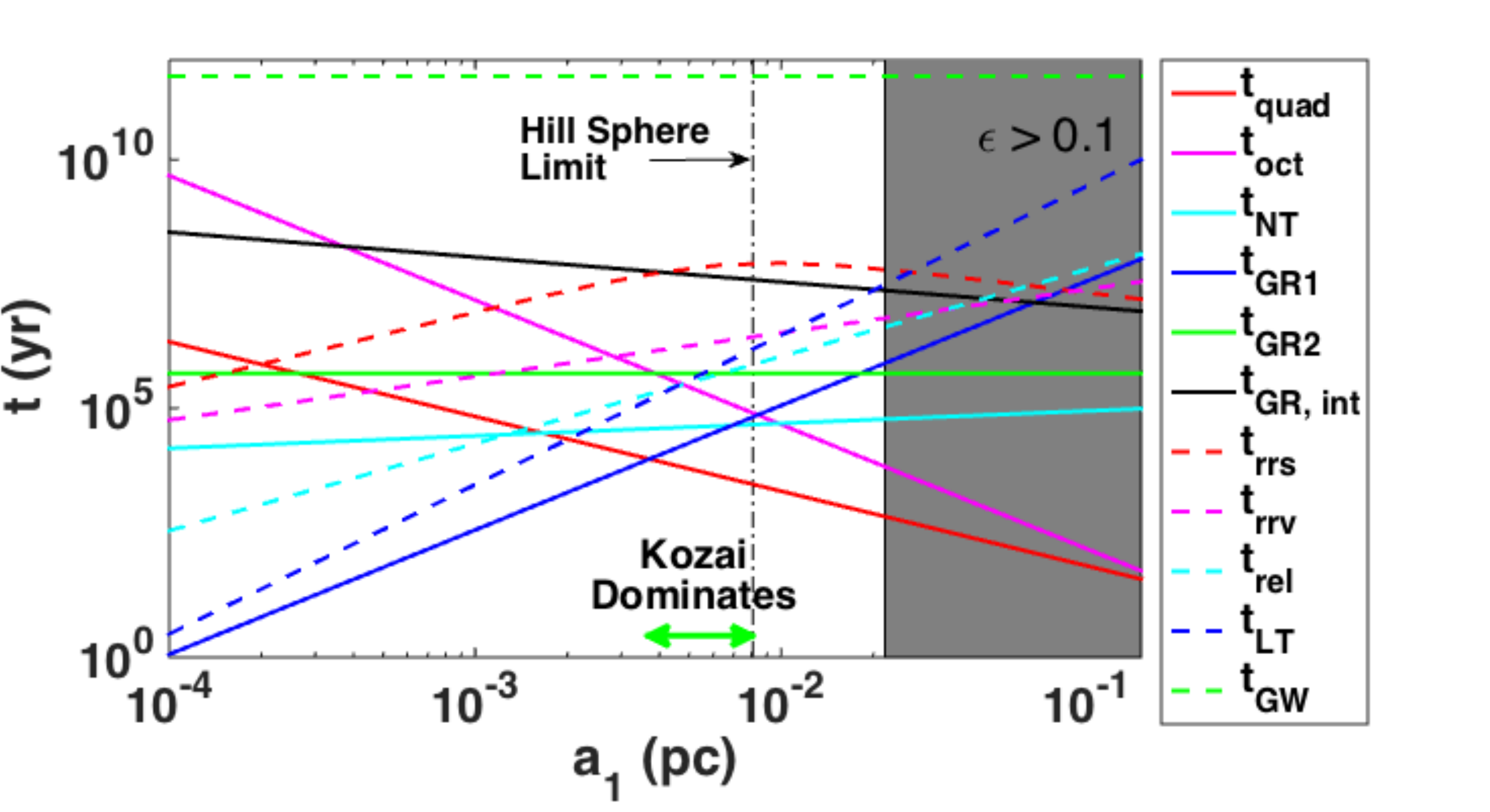}\\
\caption{\label{f:timescale} The different timescales as a function of the semi major axis of the stars ($a_1$), where $e_1=2/3$, $m_0 = 10^7 \Msun$, $a_2 = 0.3$ pc, $m_1 = 1\Msun$, $e_2=0.7$. In the upper panel, $m_2 = 10^6 \Msun$, and in the lower panel, $m_2 = 10^9\Msun$. In the grey region, $\epsilon > 0.1$, the hierarchical approximation is violated. The EKL mechanism does not operate in the grey region and wherever $t_{quad}$ is not the shortest timescale. The quadrupole Kozai timescale is shorter than the other timescales for the semimajor axis range indicated by the light green arrow.}
\vspace{0.1cm}
\end{figure}

Figure \ref{f:timescale} shows the timescales for the case of a $1\,\Msun$ star orbiting a $10^7\, \Msun$ SMBH. The separation of the SMBHB is set to $0.3$ pc. The upper panel corresponds to $m_2 = 10^6\, \Msun$, and the lower panel corresponds to $m_2 = 10^9\,\Msun$. For the Lense-Thirring timescale, $s$ is set to unity. The eccentricity of the star-SMBH system, $e_1$, is assumed to be $2/3$ and $e_2$ is assumed to be $0.7$. The EKL-dominated region is larger for higher $e_2$ with fixed $a_1$ and $a_2$. Figure \ref{f:timescale} shows that the EKL mechanism is suppressed for a $10^7$--$10^6\,\Msun$ binary at all radii, but it may operate at least in a restricted range for a $10^7$--$10^9\,\Msun$ binary. Note that although the octupole timescale $t_{oct}$ is longer than some of the other secular timescales, our simulations show that the eccentricity can nevertheless reach high values provided that $t_{K}$ is the shortest timescale and $t_{oct}$ is at most moderately larger than the other timescales. Since $t_{oct}=t_{K} /\epsilon$ and $1/\epsilon \sim 10-30$, $t_{oct}$ is only moderately larger than the other timescales in most of the relevant phase space when $t_{K}$ is the shortest timescale. Thus, in the following, we identify the regions where the eccentricity may be excited using conditions \ref{i:hierarchical}--\ref{i:quadrupole} above irrespective of $t_{oct}$. Typically, the conditions on the quadrupole Kozai timescale ($t_K < t_{GR}$ and $t_K < t_{NT}$) set the lower limit for $a_1$ for a fixed $a_2$, and the hierarchical configuration $\epsilon<0.1$ and the Hill sphere limit set the upper limit on $a_1$. 

\begin{figure}
\includegraphics[width=3.in, height=2.in]{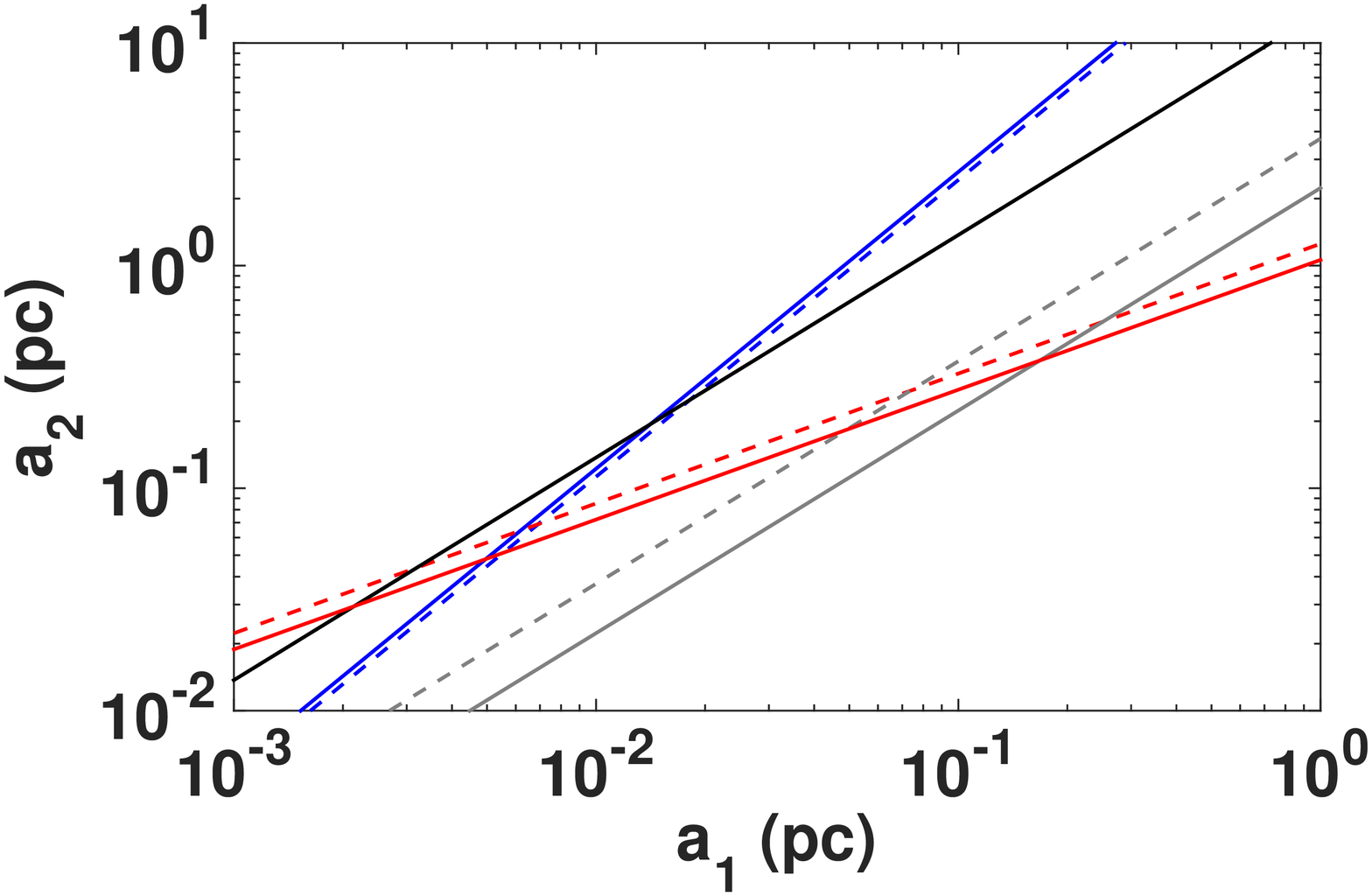}\\
\includegraphics[width=3.in, height=2.in]{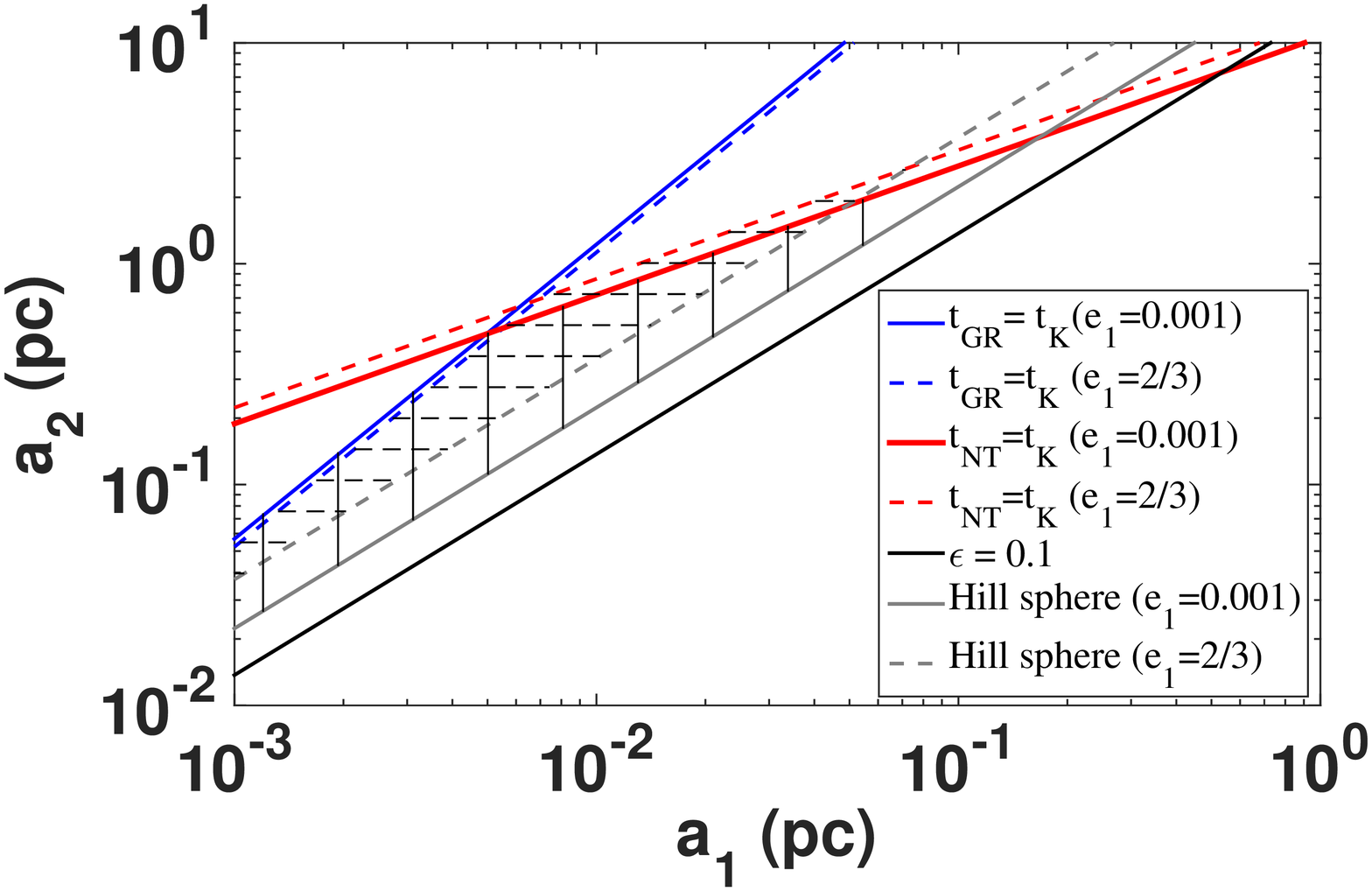}\\
\caption{\label{f:a1a2} The $a_1 - a_2$ parameter space, $m_0 = 10^7 \Msun$, $m_1 = 1\Msun$, $e_2=0.7$. In the upper panel, $m_2 = 10^6 \Msun$, and in the lower panel, $m_2 = 10^9\Msun$. The solid blue and red lines represent $e_1 = 0.001$ and the dashed blue and red lines represent $e_1 = 2/3$. Above the red or blue lines, the EKL mechanism is suppressed by the GR or the Newtonian precession. Below the black line or the grey lines, the hierarchical configuration or the Hill sphere limit is violated. The EKL mechanism is suppressed everywhere in the upper panel, and the EKL mechanism dominates in the shaded regions in the lower panel.}
\vspace{0.1cm}
\end{figure}

Next, we examine the $a_1 - a_2$ parameter space to identify the parameters where EKL dominates. We plot two examples in Figure \ref{f:a1a2}: $m_0 = 10^7\,\Msun$, $m_1 = 1\,\Msun$, $m_2 = 10^6\,\Msun$, $e_2 = 0.7$ in the upper panel, and $m_0 = 10^7\,\Msun$, $m_1 = 1\,\Msun$, $m_2 = 10^9\,\Msun$, $e_2 = 0.7$ in the lower panel. The EKL-dominated region is bigger for larger $e_2$. To test the dependence on $e_1$, we include two $e_1$ values: $e_1 = 0.001$ (solid lines) and $e_1 = 2/3$ (dashed lines), where $e_1=2/3$ corresponds to the mean value of $e_1$ in a thermal distribution. The parameter space is independent of the mass of the star as long as $m_1 \ll m_0$. The EKL-dominated region is bounded by $t_K = t_{GR}$ (blue line) and $t_K = t_{NT}$ (red line) from above and by the Hill sphere limit (grey line) and the hierarchical condition (black line) from below. In the upper panel, there is no region where the EKL mechanism dominates. In the lower panel, the region where EKL dominates is shaded with horizontal dashed lines for $e_1 = 2/3$ and it is shaded with vertical solid 
lines for $e_1 = 0.001$. 

\begin{figure*} 
\includegraphics[width=6.8in, height=4.in]{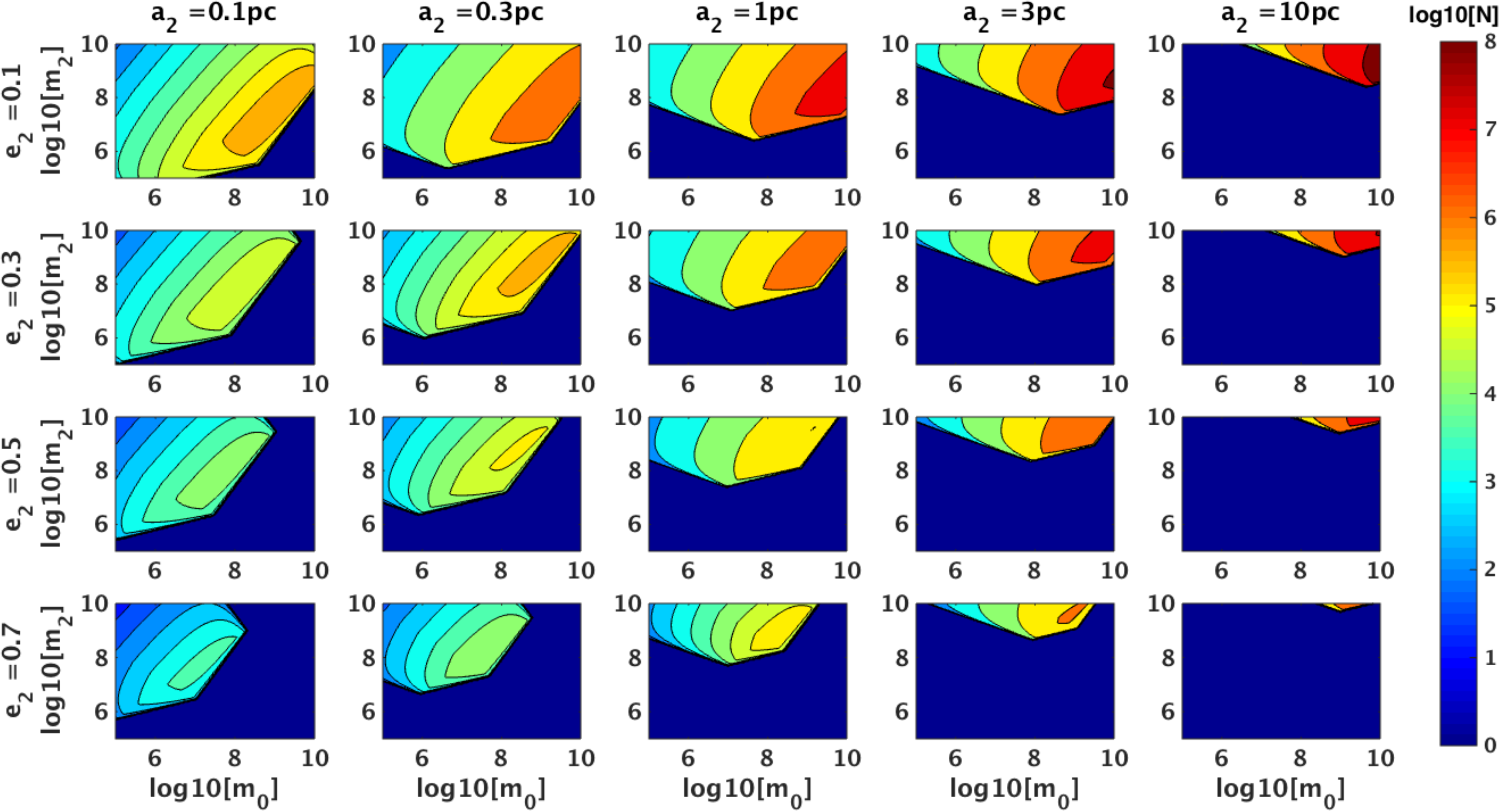}
\caption{\label{f:KN} The number of stars (N) influenced by the EKL mechanism, assuming a stellar density distribution in equation (\ref{e:rho}), and that the stellar mass is negligible and $e_1=2/3$ (the mean eccentricity assuming a thermal distribution). We determine the range of stellar semimajor axis $a_1$ where the EKL mechanism operates for a fixed set of SMBH masses, $m_0$, $m_2$, and outer orbital parameters, $e_2$ and $a_2$. Plotting the corresponding number of stars as a function of $m_0$ and $m_2$ for an array of $e_2$ and $a_2$ as shown, captures a large parameter space. The EKL mechanism affects a large number of stars over a wide range of SMBH binary parameters when $a_2\lesssim 3$ pc.
}
\vspace{0.1cm}
\end{figure*}

We calculate the number of stars affected by the EKL mechanism for the particular stellar density distribution around the inner SMBH (equation (\ref{e:rho})). In Figure \ref{f:KN}, we consider the parameter space of different $m_0$, $m_2$, $a_2$, $e_2$ and show the number of stars in the range of $a_1$ where all criteria are satisfied for the EKL mechanism to operate. Each panel shows the parameter plane of $m_0$ and $m_2$ (assuming $m_1 \ll m_0$), $a_2$ is varied in different columns of panels from 0.1 to 10 pc, and $e_2$ is varied in the different rows from 0.1 to 0.7. We set the stellar eccentricity to $e_1=2/3$ in all panels, the mean eccentricity for an isotropic thermal distribution. There is no systematic change in the number of stars affected by the EKL versus $e_1$. When $e_1=0.001$, the numbers typically increase to roughly twice the numbers of $e_1=2/3$, since the maximum $a_1$ allowed due to the Hill sphere criterion becomes larger. When $e_1=0.999$, the parameter region where stars can be affected in the $m_0 - m_2$ plane increases, since the Newtonian precession timescale increases, while the changes in the numbers depend on the specific $m_0-m_2$ configurations. In regions where the EKL mechanism is important, approximately $10^{5-6}$ stars are affected. Thus, the EKL mechanism may significantly contribute to the tidal disruption events. Note that the EKL mechanism is more likely to be suppressed for stars orbiting around the more massive SMBH. However, for parameters where the EKL mechanism is not suppressed everywhere around the more massive inner SMBH, the total number of stars affected by EKL may be higher for stars orbiting the more massive SMBH than for those orbiting the less massive SMBH. 

\subsection{Equations of motion}\label{s:EOM}
As shown in the previous section, GR and NT precessions represent important limitations for the EKL mechanism. In this section, we review the equations of motion which govern the long-term evolution of stars due to the EKL mechanism, GR and NT precessions, and tidal effects adopted from \citet{Naoz13a,Naoz13b} and \citet{Tremaine05}. 
We use the Delaunay's elements, which provide a convenient dynamical description of hierarchical three-body systems. The coordinates are the mean anomalies, $l_1$ and $l_2$, the arguments of periastron, $g_1$ and $g_2$, and the longitude of nodes, $h_1$ and $h_2$. Their conjugate momenta are \\
\begin{align}
L_1 &= \frac{m_0m_1}{m_0+m_1}\sqrt{G(m_0+m_1)a_1} \\
L_2 &= \frac{m_2(m_0+m_1)}{m_0+m_1+m_2}\sqrt{G(m_0+m_1+m_2)a_2} \nonumber \\
G_1 &= L_1\sqrt{1-e_1^2} ,
G_2 = L_2 \sqrt{1-e_2^2}\\
H_1 &= G_1 \cos{i_1},~~~ 
H_2 = G_2 \cos{i_2}  ,
\end{align}
where $i$ denotes the inclination relative to the total angular momentum of the three-body system and $G$ without subscript is the gravitational constant. 
To leading order, the two binaries follow independent Keplerian orbits where $l_j$ are rapidly varying and $L_j$, $G_j$, $H_j$, $g_j$, and $h_j$ are conserved for $j\in\{1,2\}$. These quantities are slowly varying over longer timescales due to the superposition of the perturbations: the EKL mechanism, GR and NT precessions, and tidal effects, discussed next. 

\subsubsection{Eccentric Kozai-Lidov Mechanism}
The equations of motion for the EKL mechanism may be derived using the double averaged Hamiltonian (i.e. averaged over the rapidly varying $l_1$ and $l_2$ elements). We go beyond the analyses of \citet{Chen11} and \citet{Wegg11}, who considered only the quadrupole (${\cal O} (a_1/a_2)^2$) Kozai-Lidov mechanism, where the z-component of angular momentum is constant. This assumption does not hold when the orbit of the SMBHB is eccentric, and one needs to include the octupole order terms (${\cal O} (a_1/a_2)^3$) \citep[e.g.][]{Naoz13a}. The Hamiltonian can be decomposed as
\begin{align}
{\cal H}_{Kozai, quad} &= C_2\{(2+3e_1^2)(3\cos^2{i_{tot}}-1) \nonumber \\
                    &+15e_1^2\sin^2{i_{tot}}\cos{2g_1}\} \\
{\cal H}_{Kozai, oct} &= \frac{15}{4}\epsilon_M e_1 C_2\{A\cos{\phi} 
						+10\cos{i_{tot}}\sin^2{i_{tot}} \nonumber \\
						&\times (1-e_1^2)\sin{g_1}\sin{g_2} \} ,
\end{align}
where 
\begin{align}
\epsilon_M &= \frac{m_0-m_1}{m_0+m_1} \epsilon \\
C_2 &= \frac{G^2}{16}\frac{(m_0+m_1)^7}{(m_0+m_1+m_2)^3}\frac{m_2^7}{(m_0m_1)^3}\frac{L_1^4}{L_2^3G_2^3}\\
A&=4+3e_1^2-\frac{5}{2}B\sin^2{i_{tot}}\\
B&=2+5e_1^2-7e_1^2\cos{2g_1} \\
\cos{\phi} &= -\cos{g_1}\cos{g_2}-\cos{i_{tot}}\sin{g_1}\sin{g_2}
\end{align}
The equations of motion for the EKL mechanism are given by Hamilton's equations \citep[eqn (A26-35) in ][]{Naoz13a}.

\subsubsection{GR effects}
Next, we consider the leading order (first Post-Newtonian, 1PN) effects of GR.  We follow \citet{Naoz13b}, who derived the double averaged 1PN Hamiltonian to the octupole $({\cal O}(a_1 / a_2)^3)$ order. The Hamiltonian consists of four terms: ${\cal H}_{a_1}$, ${\cal H}_{a_2}$, ${\cal H}_{a_1a_2}$, ${\cal H}_{int}$ \citep{Naoz13b}. Here ${\cal H}_{a_1a_2}$ does not contribute to the dynamical evolution, and the long-term effect of ${\cal H}_{int}$ is typically negligible, as its timescale is longer than that of the Kozai timescale and the GR precession of the inner and outer orbit as long as the star stays within the Hill sphere of the inner SMBH. Thus, we only consider the effects of ${\cal H}_{a_1}$ and ${\cal H}_{a_2}$ which cause the GR precession of the arguments of periapsides,
\begin{align}
\frac{dg_1}{dt} \Big|_{1PN, a_1} &=-\frac{3G^{3/2}(m_0+m_1)^{3/2}}{a_1^{5/2}c^2(1-e_1^2)}\,, \\
\frac{dg_2}{dt} \Big|_{1PN, a_2} &=-\frac{3G^{3/2}(m_0+m_1+m_2)^{3/2}}{a_2^{5/2}c^2(1-e_2^2)}\,.
\end{align}
Given that we neglect ${\cal H}_{int}$, and higher order Post-Newtonian corrections such as Lense-Thirring precession and gravitational radiation, the other conserved quantities, $L_j$, $G_j$, $H_j$, $h_j$, are not effected for $j\in\{1,2\}$.

\subsubsection{NT precession}
The Newtonian potential of a spherical stellar cusp causes apsidal precession at the rate \citep{Tremaine05}:
\begin{equation}
\label{e:NTpre}
\dot{g}_{1, NT} = \frac{(1-e_1^2)^{1/2}}{(Gm_0/a_1^3)^{1/2}a_1e_1}\frac{d\Phi_*}{dr}\cos{\psi} ,
\end{equation}
where $\Phi_*$ is the stellar potential, $r$ is the distance to the central SMBH and $\psi$ is the true anomaly of the inner orbit. The averaged precession rate of $g_1$ due to Newtonian precession is expressed below:
\begin{equation}
\dot{g}_{1, NT} = \frac{(Gm_0/a_1^3)^{1/2}}{\pi m_0 e_1}\int_0^\pi d\psi \, M_*(r) \cos{\psi}, 
\end{equation}
where $M_*(r)$ is the mass of the stellar system interior to $r$ and $r\equiv r(\psi)=a_1(1 - e_1^2)/(1 + e\cos{\psi})$ from Kepler's equation. Explicit analytic expressions for the apsidal precession rate are given in Appendix A of \citet{Kocsis14}.

\subsubsection{Tidal dissipation}
To investigate if tides can suppress eccentricity excitation, we consider the ``equilibrium tide'' with constant time lag to calculate the inner binary's orbital evolution when the pericenter distance is larger than $2 R_t$. Similarly to \citet{Naoz12b} and \citet{Naoz14F}, we include the differential equation governing the orbital evolution following \citet{Eggleton98, Eggleton01} and \citet{Fabrycky07}. For the star, we assume the viscous timescale is 10 yr, which corresponds to the quality factor \citep{Goldreich66} $Q\sim 10^5$ for a 10 day orbit (or $Q\sim 4\times10^8$ for a 100 year orbit).

\begin{figure}
\includegraphics[width=3.in, height=2.7in]{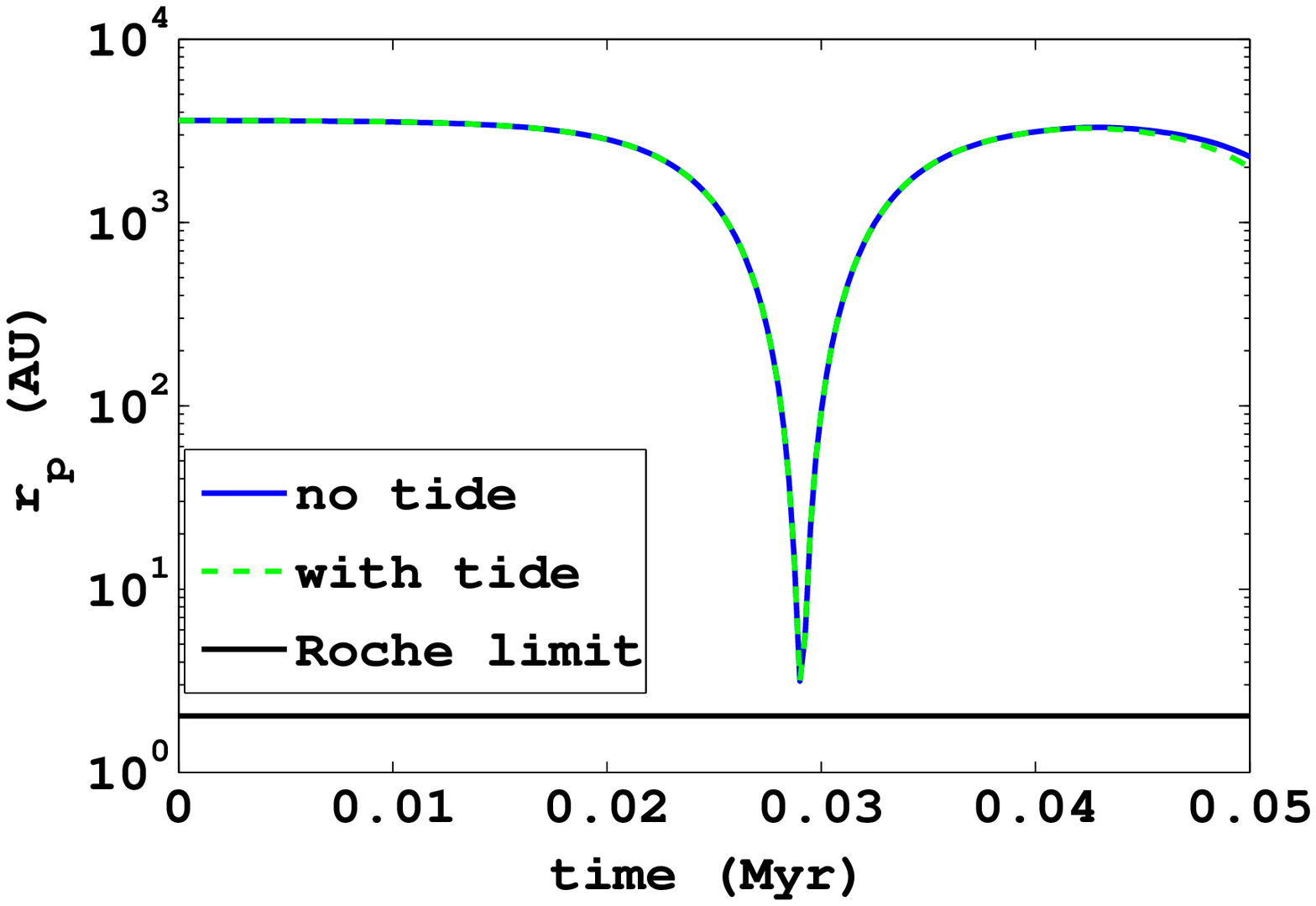}\\
\includegraphics[width=3.in, height=2.7in]{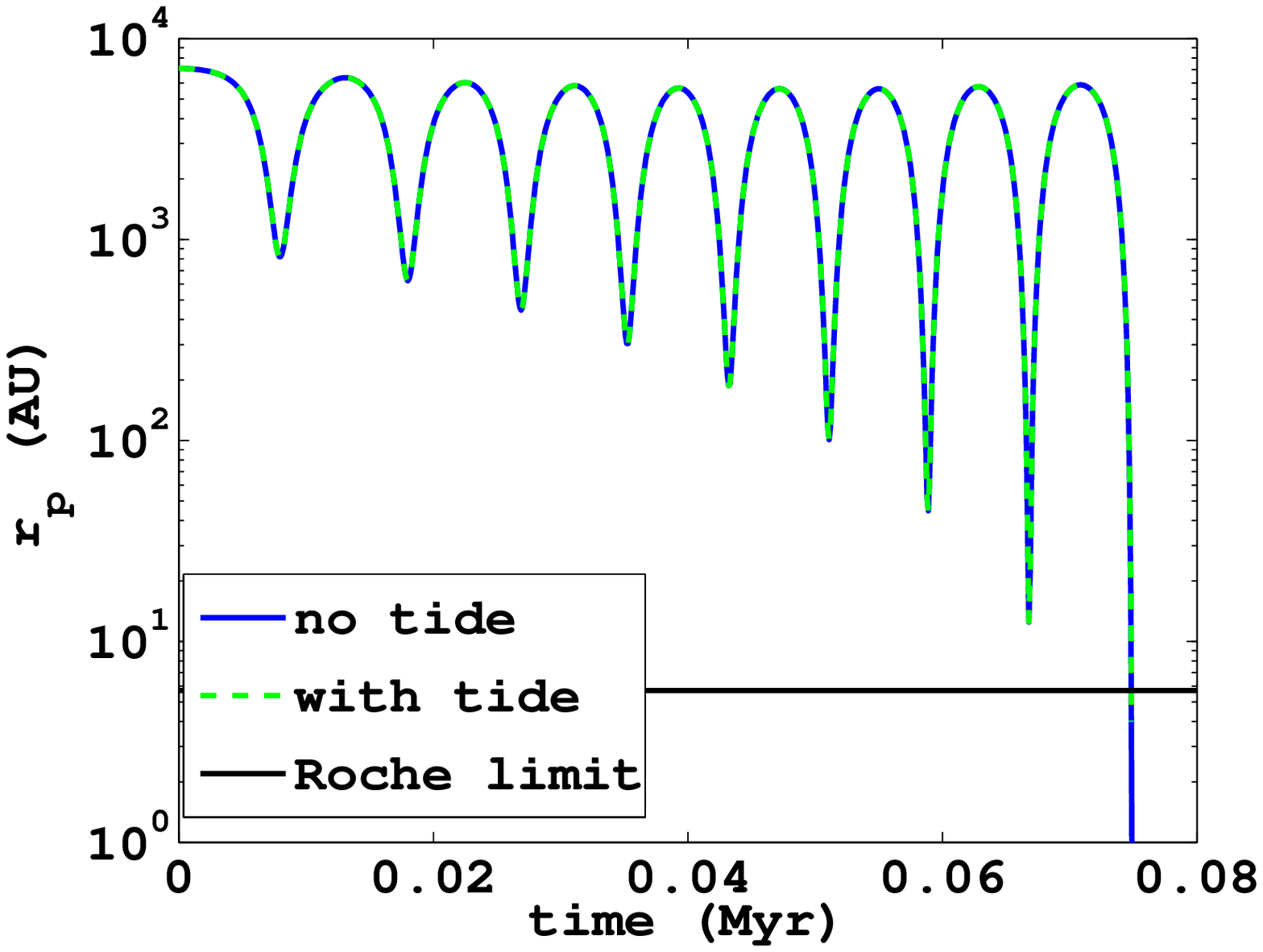}\\
\caption{\label{f: NS} Comparison of the runs with tidal effects and the runs with no tidal effects. The dashed green line indicates the case with tidal effects and the blue lines indicates the case without tidal effects. The two lines are nearly identical, suggesting that tidal effects are negligible in these runs. The upper panel shows a case when a 1 $M_{\odot}$ star orbits around a $10^7 M_{\odot}$ SMBH with $a_1 = 0.017$ pc and $e_1 = 0.001$, and is perturbed by a $10^9 M_{\odot}$ outer SMBH with $a_2 = 1$ pc. The lower panel shows a case when a 10 $M_{\odot}$ star orbits around a $10^7 M_{\odot}$ SMBH with $a_1 = 0.035$ pc and $e_1 = 0.01$, and is perturbed by a $10^9 M_{\odot}$ outer SMBH with $a_2 = 1$ pc, $e_2 = 0.7$. We used the constant time lag prescription for the tides, and the quality factor $Q$ was set to $\sim10^5$ for a 10 day orbit ($Q\sim 4\times10^8$ for a 100 year orbit).}
\vspace{0.1cm}
\end{figure}

In Figure \ref{f: NS} we show a representative example of the evolution with and without tides. The effect of tides is negligible mainly because the orbital period is long and $Q$ is low. 

\section{SMBH-binary System}
\label{s:TDR}
Requiring the criteria listed in \textsection \ref{s:Cot}, the minimum and the maximum distance of the star affected by the EKL mechanism from the inner SMBH can be calculated. However, not all stars in this region will be disrupted, since the excitation of the eccentricity depends sensitively on the orbital orientation, and the parameter region where the eccentricity can be excited is complicated \citep{Li14b}. In addition, when the Kozai timescale is only slightly smaller than the GR or the NT timescale (with $t_K$ still being the smallest), the evolution of the inner orbit is complex. For instance, the eccentricity of the inner orbit can be excited in configurations where the eccentricity cannot be excited due to the Kozai-Lidov mechanism alone. This excitation may be caused by the resonances between the NT, GR or Kozai-Lidov precessions \citep{Naoz12a}. 

We consider the following illustrative example: $m_0 = 10^7 M_{\odot}$, $m_2 = 10^8 M_{\odot}$, $a_2=0.5$ pc, $e_2 = 0.5$. We adopt the isotropic stellar distribution function of equation (\ref{e:rho}), assuming the stars have a solar mass, and that the eccentricity distribution is thermal ($dN/de = 2e$). We run large Monte-Carlo simulations, integrating the equations presented in \textsection \ref{s:Method}, where the equations of motion for the EKL mechanism are given by Hamilton's equations \citep[eqs. (A26-35) in ][]{Naoz13a}, and $\dot{g}_1 = \dot{g}_{1, EKL}+\dot{g}_{1, GR}+\dot{g}_{1, NT}$, $\dot{g}_2 = \dot{g}_{2, EKL}+\dot{g}_{2, GR}$. We distinguish three outcomes for the EKL evolution: ``TDE", ``scattered by the SMBH companion'', and ``surviving'', as explained now. 

The eccentricity of the star needs to reach very close to unity to cause tidal disruption. The tidal radius is $R_t = 5\times10^{-6}$ pc around a $10^7\Msun$ SMBH. We identify the TDE with $a_1(1-e_1) < 3 R_t$, since the stars may still be disrupted due to accumulated heating under the strong tide outside the tidal radius \citep{Li13}. Since the size of the Hill sphere of the less massive SMBH is small (i.e. $0.08$ pc in our example), the star may reach the apocenter outside the Hill sphere before disruption as the eccentricity increases. Namely, the gravitational pull of the companion SMBH ($m_2$) will be larger than $m_0$. We refer to this as a ``scattering event'' ($a_1(1+e_1) > a_2(1 - e_2)(m_0/(3m_2))^{1/3}$). Note that the secular approximation is no longer valid for the scattering events. Three-body integrations of the dynamical evolution of scattering events show that they may either lead to an exchange interaction, where the star is captured by the outer SMBH, they may cause the ejection of the star 
producing a hyper-velocity star \citep{Samsing14, Guillochon14}, or they may be tidally disrupted. The scattering events resulting in a capture may systematically increase the eccentricity distribution of stars orbiting the companion SMBH. For the third category, we label the stars neither disrupted nor scattered by the companion after 1 Gyr as ``survivors''.

\begin{figure}
\includegraphics[width=3.6in, height=2.4in]{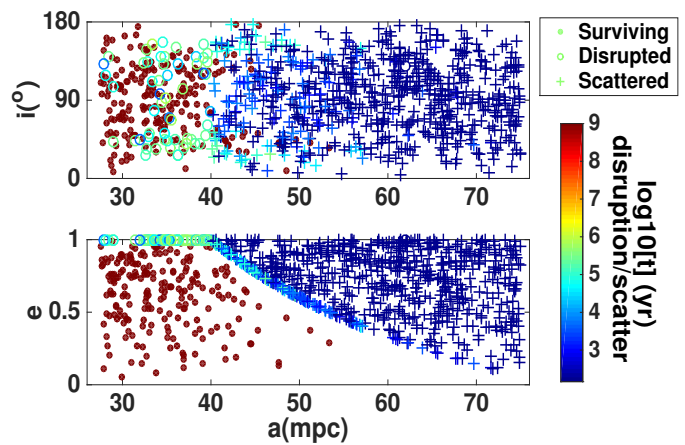}
\caption{\label{f: aei} The outcome of the evolution around a SMBH binary with $m_0 = 10^7 M_{\odot}$, $m_2 = 10^8 M_{\odot}$, $a_2=0.5$ pc, $e_2 = 0.5$. We plot the final $i_1$ versus $a_1$ and $e_1$ versus $a_1$ for stars that survived, were disrupted, or were scattered in the simulation after $1$ Gyr. The color code indicates the time when the star is disrupted or is scattered. Out of the 1,000 stars between $a_1 = 0.0275$ pc and $0.075$ pc, 57 are disrupted, and 726 are scattered by the outer black hole. The number of stars in this range according to the distribution of equation (\ref{e:rho}) is $\sim 10^5$ (assuming the stars are 1 solar mass). This suggests that the tidal disruption rate is $\sim 10^{-2}/\rm{yr}$ in the first $\sim 0.5$ Myr for the less massive black hole.}
\vspace{0.1cm}
\end{figure}

Figure \ref{f: aei} shows the results of the numerical simulation in the final $a_1 - i$ and $a_1 - e_1$ planes. We use open circles to mark stars that underwent TDEs, crosses for stars that were scattered by the companion, and full circles for stars that survived. The disruption/scattering time is color coded, and it indicates that most of the disruption events occur within  $\sim0.5$ Myr. This corresponds to the octupole Kozai timescale, which is roughly $0.2-2$ Myr for these systems at $a_1 = 0.03-0.08$ pc. Out of all 1,000 stars between $a_1 = 0.0275$ pc and $0.075$ pc, 57 are disrupted, and 726 are scattered by the outer black hole. According to the stellar density distribution in equation (\ref{e:rho}), there are $\sim 10^5$ stars in this semi-major axis range. Normalized by the total number of stars in this semi-major axis range, it indicates that the tidal disruption rate is $\sim 10^{-2}/\rm{yr}$ in the first $\sim0.5$ Myr for the less massive black hole due to EKL, while $\sim7\times10^4$ stars undergo 
scattering events by the outer SMBH. 

\begin{figure}
\includegraphics[width=3.2in, height=1.9in]{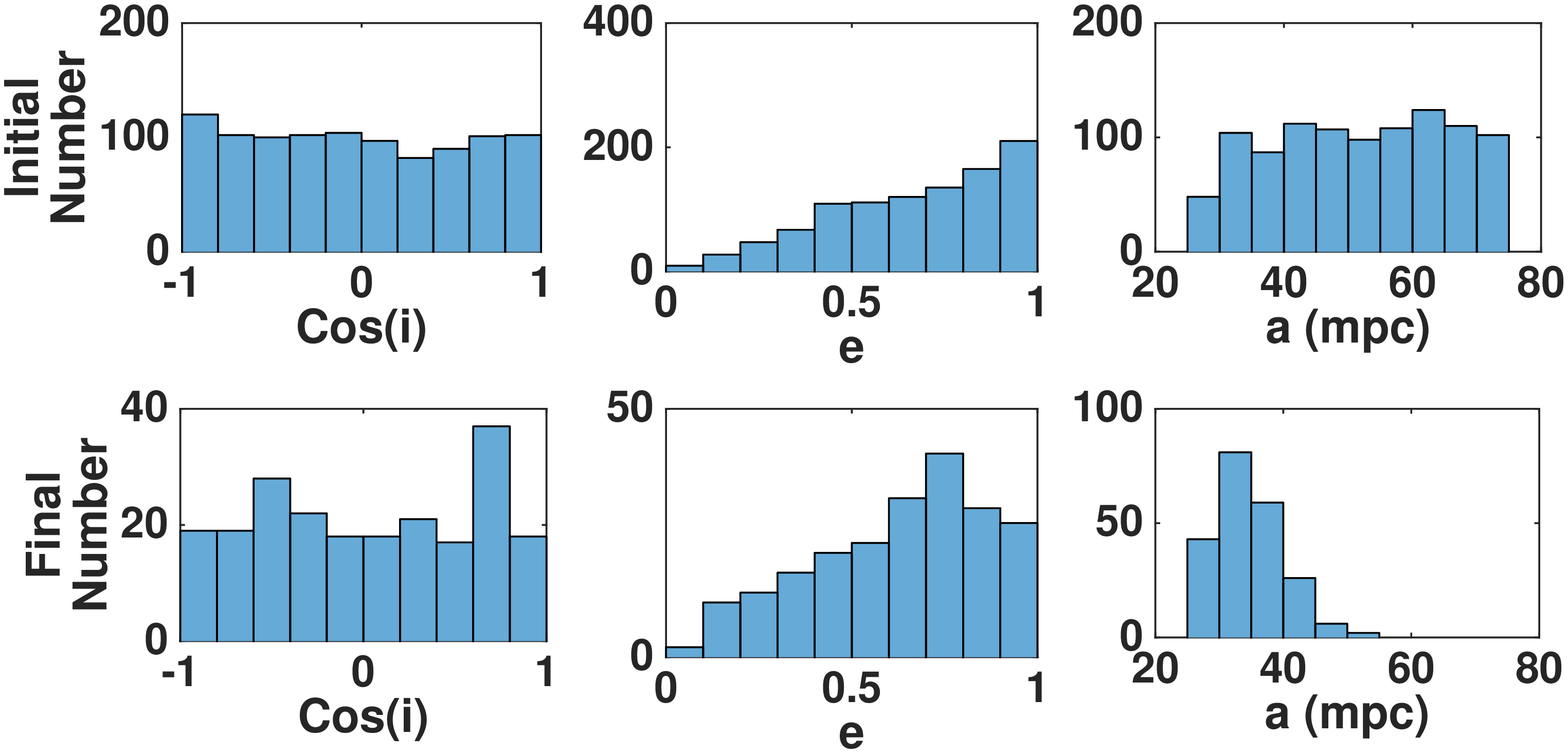}
\caption{\label{f: hist} The initial distribution and the final distribution of the stars after $1$ Gyr in our illustrative example shown in Figure~\ref{f: aei}. The final distribution represent the surviving stars.}
\vspace{0.1cm}
\end{figure}

\begin{figure}
\includegraphics[width=3.in, height=2.5in]{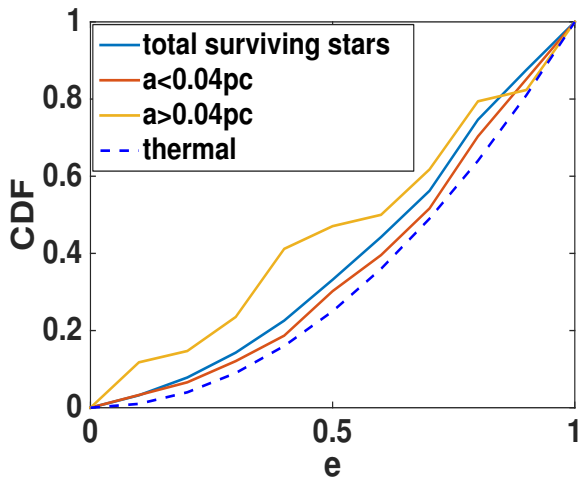}
\caption{\label{f: cdf} The final cumulative distribution of the eccentricity of stars in our illustrative example for $m_0 = 10^7 M_{\odot}$, $m_2 = 10^8 M_{\odot}$ separated by $0.5$ pc in an eccentric orbit with $e_2 = 0.5$. For stars at distance larger than $0.04$ pc, the final eccentricity distribution becomes shallower than that inside of $0.04$ pc.}
\vspace{0.1cm}
\end{figure}

Since the eccentricity of the stars with high inclinations are more likely to be excited, the stars with high inclinations are more vulnerable to tidal disruption, the final inclination distribution is no longer isotropic (the lower panels in Figure \ref{f: hist}) and the stars around the SMBH form a torus-like configuration (see \citet{Naoz14} for similar results). The stars with larger semi major axis have higher probability to be scattered when their eccentricity become excited due to the EKL mechanism, and thus the final distribution of stars surrounding the less massive black hole will be truncated at a larger semimajor axis. In addition, the distribution of the eccentricity of the surviving stars shows deviations from thermal distribution with a suppression of very eccentric stars (as expected since they get scattered by $m_2$ more easily, and their eccentricity can be excited more easily at a lower inclination \citep{Li14a}). Furthermore, as shown in Figure \ref{f: cdf}, the stars that are closer to $m_0$ ($\lesssim 0.04$ pc) have an eccentricity distribution closer to thermal. The stars that are closer to $m_2$ ($\gtrsim 0.04$ pc) have systematically smaller eccentricities. The thermal distribution for closely separated stars ($\lesssim 0.04$ pc) is similar to the observed S stars in the center of the Milky-Way galaxy \citep{Genzel10}, which shows a steeper slope.

\section{SMBH-IMBH system}
Let us consider next the perturbations of a SMBH on stars orbiting an intermediate mass black hole (IMBH). IMBHs may form through runaway mergers during core collapse in globular clusters \citep{2002ApJ...576..899P}. Since globular clusters sink to the galactic center through dynamical friction, and the disrupted globular cluster could contribute to most of the mass in nuclei stellar cluster for galaxies with total mass below $10^{11} \rm{M_{\odot}}$, this setup may be common in the Universe\citep{2006ApJ...641..319P, 2013ApJ...763...62A,Gnedin14}. Alternatively, IMBH may form at cosmologically early times from population III stars in galactic nuclei \citep{Madau01}, or in accretion disks around SMBHs \citep{2004ApJ...608..108G,2012MNRAS.425..460M,2014MNRAS.441..900M}. In the Milky Way center, the orbits of the S-stars are consistent with that caused by the dynamical interactions of IMBHs \citep{Merritt09}. In addition, IRS 13E may potentially host an IMBH, though its existence is controversial \citep{Maillard04, Schodel05, Fritz10}. The TDE rate has been discussed by \citet{Chen13} and \citet{Mastrobuono14}. Here, we consider the interactions of stars surrounding IMBHs in the center of galaxies with the central SMBH due to the hierarchical three body interactions, and consider the re-distribution of the stars as a result of the interaction. 

We set the IMBH mass to $10^4 M_{\odot}$ at a distance of $0.1$ pc from Sgr A$^*$ ($a_2 = 0.1pc$, $e_2 = 0.7$, $m_0 = 10^4M_{\odot}$ and $m_2 =4\times10^6M_{\odot}$). These parameters for the IMBH are allowed according to limits on the astrometric wobble of the radio image of Sgr A$^*$ \citep{Hansen03, Reid04}, the study of hypervelocity stars \citep{Yu03}, and the study of the orbits of S stars \citep{Gualandris09}. We set the distance of stars to be uniformly distributed between $0.00045$ pc and $0.0028$ pc. The tidal disruption radius for $1\, M_{\odot}$ stars is $4.89\times10^{-7}$ pc. The minimum distance is set by requiring the GR precession timescale to be longer than the Kozai timescale, and the maximum distance is set by requiring the stars to stay in the Hill sphere of the IMBH. Note that in this case the hierarchical criterion \ref{i:hierarchical} in Sec.~\ref{s:Cot}, $\epsilon<0.1$, is satisfied as long as the stars are within the IMBH's Hill sphere. We assume the distribution of the stellar eccentricity to be uniform. We take into account GR precession, NT precession and EKL at octupole order in the integration.

\begin{figure}
\includegraphics[width=3.4in, height=1.8in]{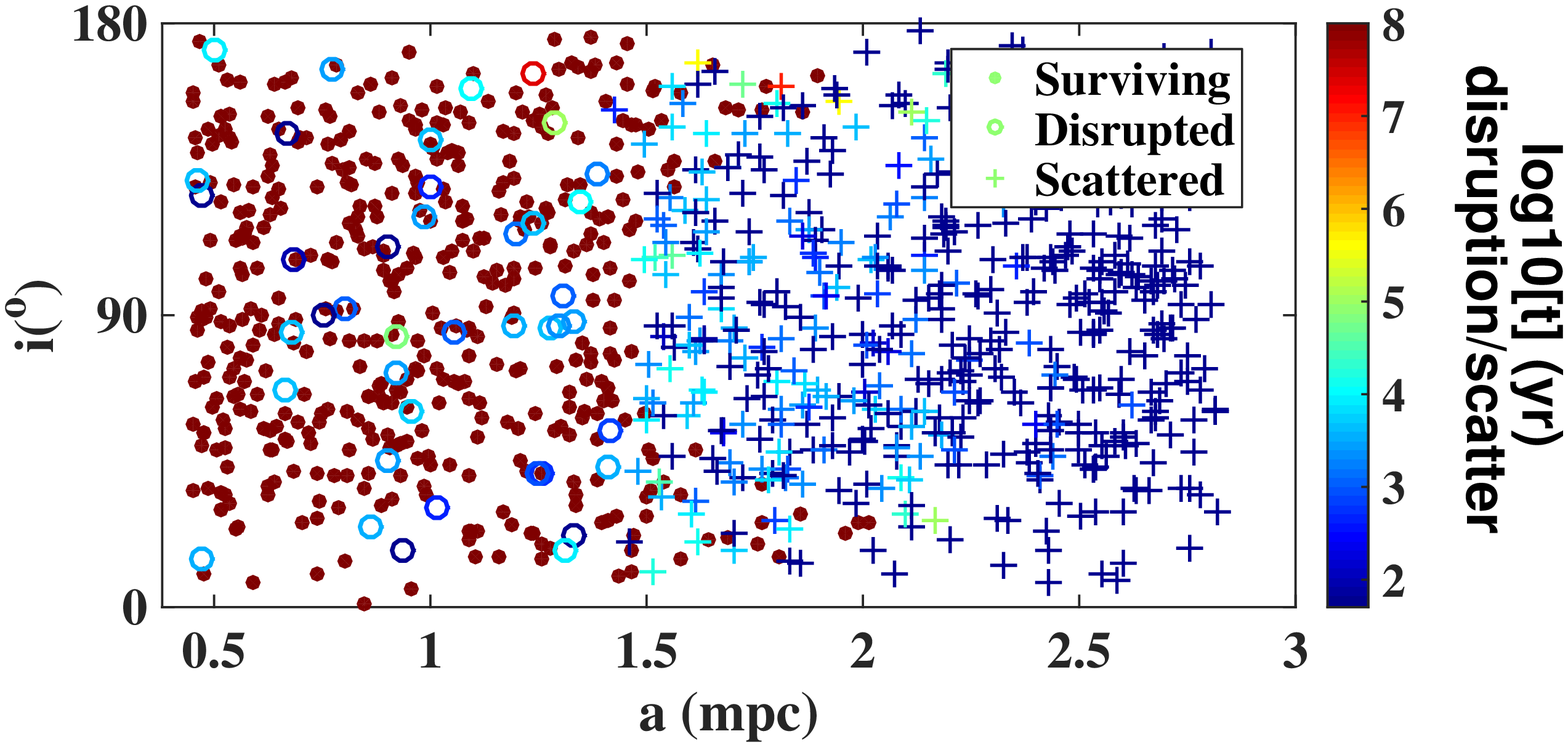}
\caption{\label{f: IMBH} The final distribution of stars surrounding a $10^4 M_{\odot}$ IMBH at a distance of $0.1$ pc from Sgr $A^*$ after $100$ Myr. The open circles represent stars that get tidally disrupted, and the crosses represent stars that get scattered. Both are colored according to the time of tidal disruption/scattering. We find that $\sim 50\%$ of the stars survived tidal disruption and scattering. The final distribution of the star has a deficiency at high inclination relative to the orbital plane of IMBH. }
\vspace{0.1cm}
\end{figure}

\begin{figure}
\includegraphics[width=3.4in, height=2.0in]{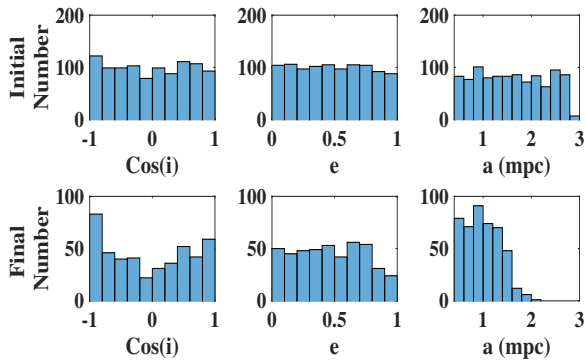}
\caption{\label{f: histIMBH} The initial distribution and the final distribution of the stars after 100Myr in our illustrative example for the IMBH, as shown in Figure \ref{f: IMBH}. }
\vspace{0.1cm}
\end{figure}

In $1,000$ runs, we find that $\sim 40$ end up in tidal disruption and $\sim 500$ are scattered as shown in Figure \ref{f: IMBH}. The tidal disruption/scattering time (color coded) is around $10^5$ yrs. As shown in Figure \ref{f: histIMBH}, we predict that the surviving stars form a torus-like configuration (similarly to the result achieved by \citet{Naoz14} for dark matter particles). The predicted distribution may be resolved if the angular resolution of the instrument is better than that corresponding to the Hill sphere around the IMBH, in this case $0.07$ arcsec. This can be achieved in near infrared by the Gemini, VLT and Keck telescopes. In addition, the EKL mechanism also produces scattering events which may be responsible for the observed hypervelocity stars. The TDE rate may reach $\sim10^{-4}/$ yr for a short $\sim 10^5$ yr duration episode after the globular cluster first approaches the galactic nucleus at a distance of $0.1$ pc, assuming there are $\sim 200$ stars in a globular cluster around an $10^4 M_{\odot}$-IMBH in the EKL-dominated region according to 
the density distribution in equation (\ref{e:rho}).


\section{Conclusions}
\label{s:conc}

SMBH binaries are natural outcomes of galaxy mergers. An SMBH binary may show an enhanced TDE rates due to the eccentric Kozai-Lidov (EKL) mechanism and chaotic three body interactions \citep{Ivanov04, Chen09, Chen11, Wegg11}. The higher tidal disruption rates may in turn serve as a flag to identify closely separated black hole binaries on subparsec scale, which are difficult to detect otherwise. 
We focused on the effect of the EKL mechanism \citep[see][]{Naoz11a, Naoz13a} on the surrounding stars in SMBHB.  This mechanism can excite the stars' eccentricity to values very close to unity \citep[e.g.,][]{Naoz13a, Naoz13b, Li14a, Li14b}. We identified the range of physical parameters where EKL is important. 

We first compared the Kozai timescale with the secular timescales of other mechanisms that may suppress EKL in galactic nuclei. These include Newtonian (NT) precession, general relativistic (GR) precession, resonant relaxation, two body relaxation, Lense-Thirring precession and orbital decay due to gravitational wave emission. We have found that for the SMBHB cases we considered, NT precession and GR precession may suppress EKL, especially when the inner SMBH is more massive than the outer SMBH (as shown in Figure \ref{f:KN}). This is consistent with the results by \citet{Naoz14} for dark matter particles around SMBH binaries, as well as the three body scattering experiments done by \citet{Chen09, Wegg11, Chen11}, who observed that the tidal disruption events were dominated by the three body chaotic interactions rather than EKL mechanism for stars surrounding the more massive black hole. However, we found that a massive outer binary allows a non-negligible
region of parameter space where the EKL mechanism may operate and lead to TDEs. We also demonstrated that tidal effects are typically negligible for the stellar orbital evolution (see Figure \ref{f: NS}). 

To illustrate the EKL effects on stars surrounding the less massive black hole, we ran 1,000 numerical experiments with different initial conditions for a star cluster surrounding a $10^7 M_{\odot}$ black hole, which is being perturbed by a $10^8 M_{\odot}$ outer black hole. We have found over $\sim 50$ out of the 1,000 runs stars are disrupted in $\sim 0.5$ Myr. Scaled with the total number of stars according to equation (\ref{e:rho}), this corresponds to a TDE rate of $10^{-2}/$yr for the first $\sim 0.5$ Myr. In contrast, \citet{Chen11} considered tidal disruption rates for stars surrounding the more massive  SMBH, using numerical three body scattering experiments. They estimated the tidal disruption rate to be as high as $0.2$ per year mainly due to three-body scattering effects\footnote{since, as we showed, the EKL is suppressed in this case}, in the first $3\times10^5$ yrs for stars surrounding a $10^7 M_{\odot}$ SMBH perturbed by an 81 times less massive outer SMBH. For the same SMBHB configuration, EKL only affects at most $\sim10^3$ stars surrounding the less massive SMBH as shown in Figure \ref{f:KN}, and affects at most $\sim10^3$ stars surrounding the more massive SMBH. Thus, EKL contributes negligibly to the total tidal disruption rate in this case, but EKL contributes significantly to the TDE rate of stars around the secondary SMBH.

The EKL mechanism also affects the stellar distribution for stars surrounding the less massive SMBH. As shown in Figure \ref{f: hist}, the survived stars within a particular range of radii are distributed in the shape of a torus \citep{Naoz14}. In addition, a large number of stars orbiting the less massive black hole will be scattered by the outer black hole following the EKL-induced eccentricity increase. In our illustrative example, $\sim 670$ out of $1000$ stars are eventually transferred to an orbit around the outer, more massive SMBH. This may produce hyper-velocity stars \citep{Guillochon14}. 

Finally, we studied the tidal disruption of stars by an IMBH during mergers of globular clusters with galactic nuclei. For an IMBH of mass $10^4\,M_{\odot}$ at a distance of $0.1$ pc from Sgr A$^*$, 4\% of stars get disrupted within the relevant distance range around the IMBH, and $\sim50\%$ get scattered within $10^5$ yrs. This yields a temporary tidal disruption rate of $\sim10^{-4}/$yr. Some of the scattering events may produce hypervelocity stars or additional TDEs. The EKL mechanism produces a torus-like stellar distribution for the surviving stars, which may be resolved by the Gemini, VLT and Keck telescopes in near infrared.
Further investigations of this process using numerical scattering experiments would be a worthwhile in the future.


\section*{Acknowledgments}
This work was supported in part by NSF grant AST-1312034. BK was supported in part by the W.M. Keck Foundation Fund of the Institute for Advanced Study and NASA grants NNX11AF29G and NNX14AM24G. The numerical calculations were
performed at the Harvard-Smithsonian Center for Astrophysics, the Institute for Theory and Computation, on Harvard Odyssey cluster.
\bibliography{msref2.bib}

\end{document}